\documentclass[prl,aps,epsf,amsfonts,floats,twocolumn,amssymb,amsmath,groupedaddress,showpacs,floatfix,nofootinbib,superscriptaddress]{revtex4}

\usepackage[]{latexsym}
\usepackage{bm}
\usepackage{subeqnarray}

\usepackage{graphicx,amssymb,amsmath}
\usepackage[english]{babel}
\usepackage{graphics}                 
\usepackage{color}                    
\usepackage{hyperref}
\usepackage{braket}

\usepackage{amsfonts}
\usepackage{bm}

\def\lf{\left}

\def\non{\nonumber}

\def\ri{\right}
\def\al{\alpha}

\def\1{{_{1}}}\def\2{{_{2}}}

\def\noHe0{:\;\!\!\;\!\!:H_e(0):\;\!\!\;\!\!:}
\def\noHm0{:\;\!\!\;\!\!:H_\mu(0):\;\!\!\;\!\!:}

\def\lf{\left}

\def\non{\nonumber}

\def\ri{\right}

\def\al{\alpha}

\def\1{{_{1}}}\def\2{{_{2}}}

\begin{document}

\title{Geometric phase of neutrinos: differences between Dirac and Majorana neutrinos}

\author{A. Capolupo}
\affiliation{Dipartimento di Fisica "E.R. Caianiello" Universit\'a di Salerno,  and INFN - Gruppo Collegato di Salerno, Italy}
\author{S.M. Giampaolo}
\affiliation{International Institute of Physics, Universidade Federal do Rio Grande do Norte, 59078-400 Natal-RN, Brazil}
\author{B. C. Hiesmayr}
\affiliation{Fakultat fur Physik, Universitat Wien, Boltzmanngasse 5, 1090 Vienna, Austria.}
\author{G. Vitiello}
\affiliation{Dipartimento di Fisica "E.R. Caianiello" Universit\'a di Salerno,  and INFN - Gruppo Collegato di Salerno, Italy}

\date{\today}
\def\be{\begin{equation}}
\def\ee{\end{equation}}
\def\al{\alpha}
\def\bea{\begin{eqnarray}}
\def\eea{\end{eqnarray}}

\begin{abstract}

We analize the non-cyclic geometric phase for neutrinos. We find that the geometric phase and the total phase associated to the mixing phenomenon provide a tool to distinguish between Dirac and Majorana neutrinos. Our results  hold for neutrinos propagating in vacuum and through the matter. Future experiments, based on interferometry,  could reveal the nature of neutrinos.
\end{abstract}


\maketitle


--{\em Introduction.}
The mixing and oscillation of neutrinos represent  one of the most intriguing
phenomena in particle physics. The main issues of  neutrino physics are the neutrino mass and the nature of the Dirac vs Majorana neutrino. Many experiments \cite{Nakamura1}-\cite{Nakamura6} have confirmed neutrino oscillations. Such a phenomenon implies  that the neutrino has mass.
Since  neutrinos are neutral particles, they can be Majorana (with neutrinos and antineutrinos being the same), or Dirac particles (with neutrinos and antineutrinos being different objects). The Majorana neutrinos allow processes in which the total lepton number is not conserved, as the neutrinoless double $\beta$ decay. Such processes are inhibited for Dirac neutrinos. Experiments based on the detection of the double beta decay may contribute   to determine the neutrino nature \cite{Giuliani}, however, at the moment the nature of the neutrino remains an unsolved question.

Here, we show that the study of the total and the geometric phases for neutrino propagation  in vacuum and through matter could open a new scenario in the investigation of the neutrino properties.

In the case of neutrinos travelling through a dense medium, as for example, the Sun or
the Earth, the neutrino propagation can be affected by interactions with the particles in the medium.
Therefore, the oscillation probabilities can be considerably different than the ones due to vacuum propagation.
Such an effect is called Mikhaev-Smirnov-Wolfenstein (MSW) effect  \cite{MSW1,MSW2}.
The phenomenon originates from the fact that electron--neutrinos (and antineutrinos) have different interactions with matter compared
to the two other flavour neutrinos.
This means that in the effective Hamiltonian $H$, which governs the propagation of
neutrinos in matter, there is an extra-potential term for the electron neutrinos. In the ordinary matter, due to the coherent forward scattering on electrons such an extra-potential term
is $A_e = \pm \sqrt{2} G_F n_e $, with $n_e $   the electron density in the matter. $G_F$ is the Fermi constant, and the positive (negative) sign applies to electron-neutrino (antineutrinos).

On the other hand, the interferometry and the study of geometric phases \cite{Berry:1984jv}-\cite{Pechal}
which appear in the evolution of many physical systems \cite{grafene,Capolupo:2013xza1,Capolupo:2013xza12} have attracted the attention in recent years. A geometric phase arises in the evolution of any state $| \psi \rangle$,
describing a quantum system characterized by a Hamiltonian defined on a parameter space (in the present case the relevant parameters are $\Delta m^2$ and the mixing angles).
The geometric phase  results from the geometrical properties of such a parameter space  and can be used to study the  properties of the system \cite{Capolupo:2013xza1}. Berry and Berry-like phases have been shown in experimental observations to characterize physical
 properties of specific quantum systems. In particular, geometric phases turn out to be by themselves physical observables \cite{Jones}-\cite{Capolupo:2013xza1}. Typical examples are graphene systems \cite{grafene}, superconducting nanocircuits and devices, semiconductors \cite{Capolupo:2013xza1},  etc.

Berry-like phases and noncyclic invariants associated to
particle oscillations (see, for example, Ref.\cite{Blasone:2009xk} and references
therein) have also been studied extensively.

In the present paper  we consider the two flavor neutrino mixing case (the discussion can be generalized to the three flavor mixing case). We show that different choices of the Majorana phase    $\phi$, violating the charge-parity (CP) symmetry,    in the mixing matrix lead to different values of the total and of the geometric  phases. The oscillation formulas, on the contrary, are independent from the Majorana phase and consequently cannot be used to distinguish between Majorana and Dirac neutrinos  \cite{Giunti2010,BilenkyA}.

Therefore, in order to derive the oscillation formulas for Majorana neutrinos, one can use indifferently any of  the mixing matrices, containing the Majorana phase  $\phi$,  obtained by the  rephasing the lepton charge fields in the charged current weak-interaction Lagrangian, (for details see Ref.\cite{Giunti2010}).   For example, one can consider the following mixing matrix
  \bea\label{U1}
\textit{U}_{  1} =  \left(
            \begin{array}{cc}
             \cos  \theta   &    \sin  \theta  e^{i \phi}\\
             -  \sin  \theta  &  \cos  \theta  e^{i \phi}\\
            \end{array}
          \right)
           \,,
          \eea
          or
          \bea\label{U2}
 \textit{U}_{ 2} =  \left(
            \begin{array}{cc}
             \cos  \theta   &    \sin  \theta   e^{-i \phi}\\
             -  \sin  \theta  e^{i \phi} &  \cos  \theta \\
            \end{array}
          \right)
           \,,
          \eea
and   obtain the same oscillation formulas. The probabilities of neutrino transitions are indeed invariant under the rephasing $\textit{U}_{\alpha k} \rightarrow e^{i \phi_{k}} \textit{U}_{\alpha k} $  $(\alpha = e, \mu; k =1,2)$.

The amplitude of transitions between different flavors  are instead depending on the    choice of the matrix $U$;  see   Eq.(22) of  \cite{Giunti2010}, here   reported for reader's convenience: $\langle\mu^{+}, \nu_{e}|\psi(t)\rangle = A_{\mu} \left(\textit{U}_{e 1}\textit{U}^{*}_{\mu 1} e^{- i E_{1} t} +
\textit{U}_{e 2}\textit{U}^{*}_{\mu 2}e^{- i E_{2} t}  \right)$. 
Obviously, when squaring the amplitudes, the phases disappear and the resulting oscillation formulas do not depend on the phase.

As we will show below, the total phase (the dynamical and  the geometric one)  depends on the transition amplitudes and,   in the case of transition between different flavors, it depends on the choice of the matrix $U$. In these transitions, different choices of $U$ lead thus  to different values of the  total  phases.

In our computations we consider the matrices corresponding to $\textit{U}_{1}$ and $\textit{U}_{2}$,   for the case of   neutrinos travelling through a dense medium.

We show that, by using the matrix  corresponding to $\textit{U}_{ 2}$ in Eq.(\ref{U2}), the geometric and the total phase due to the transition  between different flavors
are different for Majorana and for Dirac neutrinos.
On the other hand, by using the matrix corresponding to $\textit{U}_{  1}$, all the phases are independent from $\phi$ and Majorana neutrinos cannot be distinguished from the   Dirac ones.

A measurement of the total and geometric phase may thus determine  the choice of the mixing matrix $U$.
Similar results are obtain in the case of   propagation in the vacuum.

--{\em Majorana and Dirac neutrino.}
 A Majorana field has a physical  phase $\phi$ which violates the CP symmetry.
 In this case,
assuming for simplicity only two flavour eigenstates, $|\nu_e \rangle$, $|\nu_\mu\rangle$,  by considering the mixing matrix $\textit{U}_{ 2}$, one can write
   \bea
   \non
   |\nu_e \rangle & = & \cos \theta |\nu_1 \rangle + e^{- i  \phi} \sin \theta |\nu_2 \rangle
   \\\label{mix}
   |\nu_\mu\rangle & = & - e^{i  \phi}\sin \theta |\nu_1 \rangle + \cos \theta| \nu_2 \rangle\,,
   \eea
 where $\theta$ is the mixing angle and $|\nu_1 \rangle$, $|\nu_2\rangle$ are the eigenstates of the free Hamiltonian.
The phase $\phi$ cannot be eliminated since the Lagrangian of Majorana neutrinos
    is not invariant under $U(1)$ global transformation and the rephasing is not possible (see below).
In  this case,
only left-handed components of the neutrino  fields $ \nu
_{eL}=\frac{1+\gamma _{5}}{2}\nu _{e},$  $ \nu _{\mu
L}=\frac{1+\gamma _{5}}{2}\nu _{\mu }$ and right-handed
components of the antineutrino  fields $\nu
_{eR}^{C}=\frac{1-\gamma _{5}}{2}\nu _{e}^{C}=\left( \nu
_{eL}\right) ^{C},\qquad \nu _{\mu R}^{C}=\frac{1-\gamma
_{5}}{2}\nu _{\mu }^{C}=\left( \nu _{\mu L}\right) ^{C} $ appear in the Hamiltonian.
Here $\gamma^{5} = \frac{i}{4!}\epsilon_{\mu\nu\alpha\beta} \gamma^{\mu} \gamma^{\nu} \gamma^{\alpha} \gamma^{\beta}$, with
  $\gamma^{j}$ Dirac matrices \cite{Per}, $ \nu _{e,\mu }^{C}=C\bar{\nu }_{e,\mu }$
is the charge conjugated spinor and the  matrix $C$ satisfies the
relations: $ C^{\dagger }C = 1\,,   C\gamma _{\alpha }^{T}C^{-1} = -\gamma _{\alpha }\,,
\  C^{T} = -C \,.$
For Majorana neutrinos the interaction Hamiltonian, which does not conserve the
lepton numbers, has the  form
\begin{equation}
\label{Hsuperweak}\non
H \! \!=\!\! m_{\bar{e}e}\bar{\nu }_{eR}^{C}\nu _{eL}+m_{\bar{\mu }\mu }%
\bar{\nu }_{\mu R}^{C}\nu _{\mu L}
+
m_{\bar{\mu }e}\left( \bar{%
\nu }_{\mu R}^{C}\nu _{eL}+\bar{\nu }_{eR}^{C}\nu _{\mu L}\right)\!
+ h.c.
 \end{equation}
 where the parameters $m_{\bar{e}e}$, $m_{\bar{\mu}\mu
}$, $m_{\bar{\mu}e}$ have the dimensions of a mass.

On the contrary, for Dirac neutrino, the Lagrangian is invariant under $U(1)$ global transformation  and the
phase $\phi$ can be removed.
The Hamiltonian interaction in this case is
\begin{equation}
\label{HsuperweakDirac}\non
H = m_{\bar{e}e}\bar{\nu }_{e } \nu _{e }+m_{\bar{\mu }\mu }%
\bar{\nu }_{\mu  } \nu _{\mu }
+
m_{\bar{\mu }e}\left( \bar{%
\nu }_{\mu } \nu _{e }+\bar{\nu }_{e } \nu _{\mu  }\right)\,,
\end{equation}
where $\bar{\nu }_{\sigma }$, with $\sigma = e, \mu$ are the antineutrino fields.

The phase $\phi$ does not affect the oscillation formulas for neutrino propagating in the vacuum and through matter.
  Therefore, neglecting the dissipation  \cite{Benatti}, the oscillation formulas cannot reveal the nature of neutrinos~\cite{note}.

--{\em Neutrinos propagating through the matter.}
In the case of neutrinos propagating through the matter,
the evolution equation in the flavour basis  is   described by the Schr\"odinger equation
\bea\non
i \frac{d}{d t } \left(
                   \begin{array}{c}
                     | \nu_{e}  \rangle \\
                     | \nu_{\mu}  \rangle \\
                   \end{array}
                 \right) = \textit{H}_{f} \left(
                   \begin{array}{c}
                     | \nu_{e}  \rangle \\
                     | \nu_{\mu}  \rangle \\
                   \end{array}
                 \right)\,,
\eea
where the Wolfenstein effective Hamiltonian \cite{MSW2} for flavored neutrinos can be written as (for a formal derivation see also Ref.\cite{Bilenky1})
\bea\label{Hmat}
\textit{H}_{f} = \left( p + \dfrac{ m_{1}^{2} +  m_{2}^{2}}{4 E} + \frac{ \sqrt{2} }{2 }G_{F} \lf (n_{e}- n_{n}\ri) \right) {\bf I} +
\textit{H}_{I}.
\eea
Here, $p$ is the neutrino momentum, $m_i$ are the masses of the free fields $\nu_i$ $(i = 1, 2) $ (for relativistic neutrinos  $E_i \approx p + \frac{m_{i}^{2}}{2 E}$), $n_e$ is the electron density, $n_n$ is the neutron number density, $G_{F}$ the Fermi weak coupling constant and $H_I$ for Majorana neutrinos is \cite{Wagner}
\bea \label{H1}
\textit{H}_{I} =
 \left(
               \begin{array}{cc}
                 \frac{G_{F}n_{e}}{\sqrt{2 }} - \frac{\Delta m^{2}}{4 E} \cos 2\theta  & \frac{\Delta m^{2}}{4 E} e^{- i \phi} \sin 2\theta  \\
                 \frac{\Delta m^{2}}{4 E} e^{  i \phi} \sin 2\theta &  -\frac{G_{F}n_{e}}{\sqrt{2 }} + \frac{\Delta m^{2}}{4 E} \cos 2\theta  \\
               \end{array}
             \right)\,.
\eea

  ${\bf I}$ is the $2 \times 2$ identity matrix,  $\phi$ the CP violating phase (which can be put equal to zero in the Dirac neutrino case), $E$ the neutrino energy and $\Delta m^{2} = m^{2}_{ {2}}- m^{2}_{ {1}}$.
   The first term of \textit{H} (the one proportional to the identity matrix) is responsible only of an overall phase factor in the neutrino evolution, therefore we will neglected it in the computation of the geometric phase and focus on $\textit{H}_{I}$.

   By defining  $\Delta m_{m}^{2} = \Delta m^{2} R$, $\sin 2\theta_{m} =  \sin 2\theta /R$, with
\bea\label{R}
R = \sqrt{\left(\cos 2\theta - \frac{2 \sqrt{2 } G_{F} n_{e} E}{\Delta m^{2}}\right)^{2}+ \sin^{2} 2\theta} ~,
\eea
$\textit{H}_{I}$ in Eq.~(\ref{H1}) can be written as
  \bea\label{Hii}
\textit{H}_{I} =
 \frac{\Delta m_{m}^{2} }{4 E}\left(
               \begin{array}{cc}
                  - \cos 2\theta_{m}  &   e^{- i \phi} \sin 2\theta_{m}  \\
                   e^{  i \phi} \sin 2\theta_{m} &    \cos 2\theta_{m}  \\
               \end{array}
             \right)\,.
\eea

The eigenvalues of $\textit{H}_{I}$ are $\lambda_{\pm} = \pm \frac{\Delta m_{m}^{2} }{4 E}$. Moreover, $\textit{H}_{I}$ can be diagonalized by means of the matrix $\textit{U}_{m}$,
\bea\label{U}
\textit{U}_{m} =  \left(
            \begin{array}{cc}
             \cos  \theta_{m}  &   e^{- i \phi} \sin  \theta_{m} \\
             - e^{ i \phi}\sin  \theta_{m} &  \cos  \theta_{m} \\
            \end{array}
          \right)
           \,,
          \eea
so that $ \textit{H}_{I} = \textit{U}_{m} \textit{H}_{0} \textit{U}_{m}^{-1}$, with
$ \textit{H}_{0} =  diag (\lambda_{+}, \lambda_{-}) $.
The matrix $\textit{U}_{m}$ relates the  flavor states $|\nu_{e}(z) \rangle $ and $| \nu_{\mu}(z) \rangle$ with the energy eigenstates
$|\nu_{1}(z) \rangle $ and $| \nu_{2}(z) \rangle$. Here we consider, in natural units, the approximation, $z \approx t$, where $t$ is the neutrino propagation time and $z$ the distance traveled by neutrinos.
Explicitely one has
\bea\label{stato1}\non
|\nu_{e}(z) \rangle \!\!& = &\!\! \cos  \theta_{m} e^{ i \frac{\Delta m_{m}^{2} }{4 E} z}  |\nu_{1} \rangle
+
e^{- i \phi} \sin \theta_{m} e^{ -i \frac{\Delta m_{m}^{2} }{4 E} z}|\nu_{2} \rangle ,
\\\label{stato2}\non
|\nu_{\mu}(z) \rangle \!\!& = &\!\! - e^{ i \phi} \sin \theta_{m} e^{ i \frac{\Delta m_{m}^{2} }{4 E} z}  |\nu_{1} \rangle
+
 \cos \theta_{m} e^{ -i \frac{\Delta m_{m}^{2} }{4 E} z}|\nu_{2} \rangle ,
\\
\eea
with  $\langle \nu_{e}(z)|\nu_{e}(z) \rangle = \langle \nu_{\mu}(z)|\nu_{\mu}(z) \rangle = 1$ and
$\langle \nu_{e}(z)|\nu_{\mu}(z) \rangle = 0$.
In the case of the propagation in vacuum, $R = 1$, then
$\Delta m_{m}^{2} \rightarrow \Delta m ^{2}$, $ \theta_{m} \rightarrow \theta $
and $U_{m}$ in Eq.(\ref{U}) coincides with $U_{2} $ of Eq.(\ref{U2}).

--{\em Geometric phases for neutrinos in matter.}
In the following we consider the states given by Eq. (\ref{stato2}).
To compute the geometric and the total phases generated by the states (\ref{stato2}), we use the definition of Mukunda-Simon geometric phase \cite{Mukunda}, which generalizes
the Berry phase to the non-adiabatic and non-cyclic case.
Such a phase, derived within a kinematical approach, is associated to any open curve of unit vectors in Hilbert space.
For a quantum system whose state vector $|\psi(s)\rangle$ belongs to a curve $\Gamma$, with the real parameter $s$ such that $s \in [s_1, s_2]$, the Mukunda--Simon phase is defined as the difference between the total and the dynamic phase:
\bea\label{geofase}\nonumber
\Phi^{g } (\Gamma)\!\!& = & \!\! \Phi^{tot}_{\psi}(s) - \Phi^{dyn}_{\psi}(s)
\\
\!\!& = & \!\!
\arg \langle \psi(s_{1})| \psi(s_{2} )\rangle - \Im\int_{s_{1}}^{s_{2}}\langle \psi(s)|\dot{\psi}(s)\rangle d s
\,.
\eea
Here the dot denotes the derivative with respect to the real parameter $s$.
 In Eq.(\ref{geofase}), $\arg \langle \psi(s_{1})| \psi(s_{2} )\rangle$ represents the total phase, and $ \Im\int_{s_{1}}^{s_{2}}\langle \psi(s)|\dot{\psi}(s)\rangle d s$ is the dynamical one.

In the specific case of neutrinos, the geometric phase of electron neutrino, for an initial  state $|\nu_{e}  \rangle$, is
\bea
\Phi^{g }_{\nu_{e}}(z) \!\!& = & \!\! \Phi^{tot}_{\nu_{e}}(z) - \Phi^{dyn}_{\nu_{e}}(z)
\\ \nonumber
\!\!& = & \!\!  \arg \left[\langle \nu_{e}(0)| \nu_{e}(z)\rangle  \right] - \Im \int_{0}^{z}  \langle  \nu_{e}(z^{\prime})| \dot{\nu}_{e}(z^{\prime})\rangle  d z^{\prime}\,,
 \eea
and explicitly one obtains
\bea\non\label{fase1}
\Phi^{g }_{\nu_{e}}(z) & = & \arg \left[ \cos \left(\frac{\Delta m_{m}^{2} z}{4 E}\right) + i \cos 2\theta_{m} \sin \left(\frac{\Delta m_{m}^{2} z}{4 E}\right) \right]
\\ & - & \frac{\Delta m_{m}^{2} z}{4 E} \,\cos 2\theta_{m}  \,.
\eea

 In a similar way, the geometric phase for the muon neutrino, for an initial  state $|\nu_{\mu}  \rangle$, is given by $\Phi^{g }_{\nu_{\mu}}(z) =
\arg \left[\langle \nu_{\mu}(0)| \nu_{\mu}(z)\rangle  \right] - \Im \int_{0}^{z}  \langle  \nu_{\mu}(z^{\prime})| \dot{\nu}_{\mu}(z^{\prime})\rangle  d z^{\prime}$. One has  $\Phi^{g }_{\nu_{\mu}}(z) = - \Phi^{g }_{\nu_{e}}(z)$. Eq.(\ref{fase1}) does not depends on the $CP$ violating phase, thus, it holds both for Majorana and for Dirac neutrinos.

However, in addition to the phases $\Phi^{g }_{\nu_{e}}(z)$ and $\Phi^{g }_{\nu_{\mu}}(z)$, one also has the following phases due to the neutrino transitions between different flavors,
 \bea \nonumber\label{fasemix1}
\Phi^{g }_{\nu_{e}\rightarrow \nu_{\mu}}(z)
\!\!& = &\!\!
\arg \left[\langle \nu_{e}(0)| \nu_{\mu}(z)\rangle  \right] - \Im \int_{0}^{z}  \langle  \nu_{e}(z^{\prime})| \dot{\nu}_{\mu}(z^{\prime})\rangle  d z^{\prime}\, ,
\\
\\\nonumber \label{fasemix2}
\Phi^{g }_{\nu_{\mu}\rightarrow \nu_{e}}(z)
\!\!& = &\!\!
\arg \left[\langle \nu_{\mu}(0)| \nu_{e}(z)\rangle  \right] - \Im \int_{0}^{z}  \langle  \nu_{\mu}(z^{\prime})| \dot{\nu}_{e}(z^{\prime})\rangle  d z^{\prime}\, .
\\
\eea

\begin{figure}
\begin{picture}(300,180)(0,0)
\put(10,20){\resizebox{8.0 cm}{!}{\includegraphics{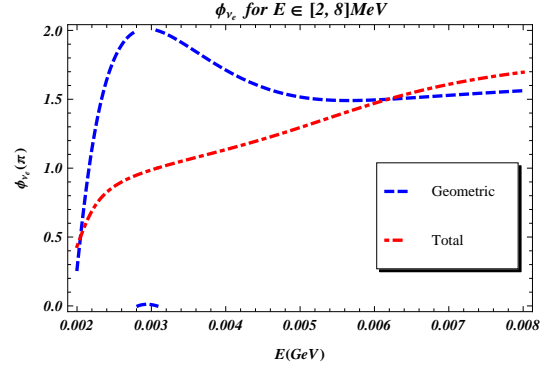}}}
\end{picture}\vspace{-1cm}
\caption{\em (Color online) Plots of the total and the geometric phases associated to the evolution of  $ \nu_{e}  $, as a function of the neutrino energy $E$, for a distance length $z = 100 km$.
  - The  red dot dashed line is the total phase; - the  blue  dashed line is the geometric phase.}
\label{pdf}
\end{figure}

 In this case, by using the states in Eq.(\ref{stato1})    for Majorana neutrinos, since $\phi$ cannot be removed, one obtains
\bea \nonumber
\Phi^{g }_{\nu_{e}\rightarrow \nu_{\mu}}(z)
 \!\!& = &\!\!
 \arg \left[\sin 2\theta_{m} \sin \left(\frac{\Delta m_{m}^{2} z}{4 E}\right)  \left(\sin \phi - i \cos \phi\right) \right]
\\
 \!\!& + &\!\!  \left(\frac{\Delta m_{m}^{2}}{4 E}\,\sin 2\theta_{m}\; \cos \phi\; \right) z\,,
 \eea
 which holds provided there is mixing ($\Delta m^2 \neq 0$ and $\theta \neq 0$). Thus, we have
 \bea
\label{fasemix1a}
\Phi^{g }_{\nu_{e}\rightarrow \nu_{\mu}}(z)   =   \frac{3\pi}{2} + \phi
+  \left(\frac{\Delta m_{m}^{2}}{4 E}\,\sin 2\theta_{m}\; \cos \phi\; \right) z\, ,
\eea
and in similar way
\bea \label{fasemix2a}
\Phi^{g }_{\nu_{\mu}\rightarrow \nu_{e}}(z)
  =   \frac{3\pi}{2} - \phi
+  \left(\frac{\Delta m_{m}^{2}}{4 E} \sin 2\theta_{m}\; \cos \phi\ \right) z\, .
\eea
 Then, $\Phi_{\nu_{e}\rightarrow \nu_{\mu}} \neq \Phi_{\nu_{\mu}\rightarrow \nu_{e}} $, which reveals  an  asymmetry between
 the transitions  ${\nu_{e}\rightarrow \nu_{\mu}} $ and $  {\nu_{\mu}\rightarrow \nu_{e}} $ due to the presence of $\phi$.

 On the contrary,
  in the case of Dirac neutrinos, the phase $\phi$ can be removed by means of $U(1)$ global transformation,
then, the phases of Eqs.(\ref{fasemix1}) and (\ref{fasemix2}) (or equivalently Eqs. (\ref{fasemix1a}) and (\ref{fasemix2a})) become equivalent
 and reduce to
 \bea
 \label{fasemixD}\non
\Phi^{g }_{\nu_{e}\rightarrow \nu_{\mu}}(z) & = & \Phi^{g }_{\nu_{\mu}\rightarrow \nu_{e}}(z)
\\
& = &
\frac{3\pi}{2}
+  \left(\frac{\Delta m_{m}^{2}}{4 E}\,\sin 2\theta_{m}\;  \right) z .
\eea

Let us now observe that the result of Eq.(\ref{fase1}) is independent on the choice of the mixing matrix.
 The same result is indeed obtained by using, for example, the mixing matrix corresponding to $U_1$ which diagonalize the interaction Hamiltonian $H_{I}^{D} $, derived by $H_{I}  $ in Eq.(\ref{Hii}) by  setting $\phi = 0$.

On the contrary, the phases defined in Eqs.(\ref{fasemix1}) and (\ref{fasemix2}) are dependent on the choice of the mixing matrix.
Indeed, the result of Eq.(\ref{fasemixD})  is obtained also for Majorana neutrinos when one considers the mixing matrix corresponding to
$U_{1}$.
We therefore conclude that the  phases $\Phi^{g }_{\nu_{e}\rightarrow \nu_{\mu}}$ and $\Phi^{g }_{\nu_{\mu}\rightarrow \nu_{e}}$
discriminate between the two matrices $U_1$ and $U_2$.

Notice that all the above phases vanish in the limit of zero neutrino masses where the mixing is absent and  Dirac and Majorana neutrinos are equivalent.

--{\em Numerical analysis.}
In our analysis, for the total and geometric phases associated with the evolution of $\nu_{e}$, we consider  the energy of neutrinos produced in nuclear reactors $E \in [2 - 8] MeV$, the earth density $n_{e } =10^{24} cm^{-3} $, $\Delta m^{2} = 7.6 \times 10^{-3}eV^{2}$ and a distance $z = 100 km$. We obtain the results reported in Fig.$1$ which could be detected in experiments like RENO \cite{Nakamura2}. Similar results can be found considering  energies of few $GeV$, which are characteristic of neutrino beams produced at particle accelerators, and distances of several hundred of km, as in long base line experiments.  For the analysis of geometric phases due to the transition between different flavors   (Eqs.(\ref{fasemix1}) and (\ref{fasemix2})), we consider energies  $E \sim 1 GeV$ and a distance $z = 300 km$, which are values similar to the ones in $T2K$ experiment. Moreover we consider $\phi = 0.3$, and the values of $n_e$ and $\Delta m^{2}$ considered above.  The plots of  the phases in Eqs.(\ref{fasemix1a}), (\ref{fasemix2a}) and (\ref{fasemixD}) are reported in Fig.$2$.

\begin{figure}
\begin{picture}(300,180)(0,0)
\put(10,20){\resizebox{8.0 cm}{!}{\includegraphics{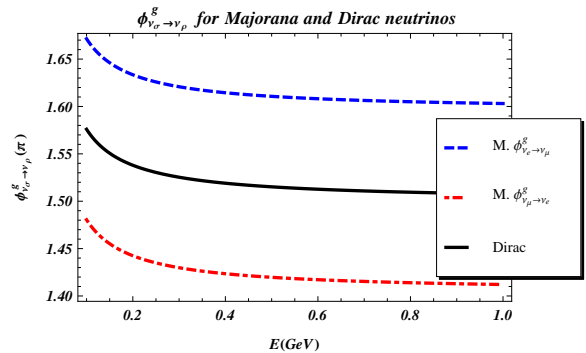}}}
\end{picture}\vspace{-1cm}
\caption{\em (Color online) Plot of the geometric phases $\Phi^{g}_{\nu_{e}\rightarrow \nu_{\mu}} $ (the blue dashed line)  and $\Phi^{g}_{\nu_{\mu}\rightarrow \nu_{e}} $ (the red dot dashed line) for Majorana neutrinos as a function of the neutrino energy $E$, for a distance length $z = 300 km$. The geometric phases $\Phi^{g}_{\nu_{e}\rightarrow \nu_{\mu}} =\Phi^{g}_{\nu_{\mu}\rightarrow \nu_{e}} $ for Dirac neutrinos  is represented by the black solid line.}
\label{pdf}
\end{figure}

--{\em Conclusions.}
We have shown that  the total and geometric phase,  due to a transition between different neutrino flavors,
  take different values depending on the representation of the mixing matrix and on the nature of neutrinos.
Measurement of different values for $\Phi^{g}_{\nu_{e}\rightarrow \nu_{\mu}}$ and $\Phi^{g}_{\nu_{\mu}\rightarrow \nu_{e}} $
could allow the detection of the Majorana phase $\phi$.
Such a measurement   implies that the mixing matrix to be used is of type $U_2$, which then removes the ambiguity in the use of $U_1$ and $U_2$. On the other hand, if the measurement of the   geometric phase leads to
   $\Phi^{g}_{\nu_{e}\rightarrow \nu_{\mu}} = \Phi^{g}_{\nu_{\mu}\rightarrow \nu_{e}} $,
 then the ambiguity between $U_1$ and $U_2$ remains and nothing can be said on the nature (Dirac or Majorana) of neutrinos.
The  phases above analyzed are in principle detectable, and next long baseline neutrino oscillation experiment  like  $T2K$ \cite{T2K}, or short base line experiments like  RENO \cite{Nakamura2}, could reveal the
 correct mixing matrix for Majorana neutrinos and the nature of the neutrinos by means of interferometric analysis.

  In a forthcoming paper we will consider the case of Dirac and Majorana neutrinos propagating through a magnetic field.

The theoretical aspects of particle mixing have been studied extensively in the
contexts of quantum mechanics (QM) \cite{Kabir}–-\cite{fidecaro} and of quantum
field theory (QFT) \cite{Blasone:1998hf}–-\cite{Blasone:2006jx} where corrections to the QM
oscillation formulas in vacuum have been derived (see also Refs. \cite{Capolupo:2006et}–-\cite{Capolupo:2010ek}).
In the discussion here presented  these corrections can be safely neglected  \cite{CapolupoPLB2004}.

-{\em Acknowledgements}
A.C. and G.V. acknowledge partial financial support from MIUR and INFN,
S.M.G. financial support from the Ministry of Science, Technology and Innovation of Brazil and
B.C.H. Austrian Science Fund (FWFP26783).


\begin{thebibliography}{99}
\bibitem{note}
 Note that dissipative effects in neutrino oscillation in matter can be
mostly neglected since they are  of the order of $E^{2}/M_{P}$, with $E$, neutrino energy and $M_{P}$ Plank Mass.
For more details see Ref.~\cite{Benatti}.
\end{thebibliography}
\end{document}